\documentclass[aps,reprint,floatfix]{revtex4-1}
\usepackage{epsfig}
\usepackage{amsthm,amsmath}
\usepackage{color}
\usepackage{amsfonts}

\begin{document}
\title{{Electron-nucleus scalar-pseudoscalar interaction in PbF: Z-vector study in the relativistic coupled-cluster framework}}
\author{
Sudip Sasmal$^{a}$,
Kaushik Talukdar$^{b}$,
Malaya K. Nayak$^{c}$,
Nayana Vaval$^{a}$
and Sourav Pal$^{b}$
\\\vspace{6pt}
$^{a}${\em{Electronic Structure Theory Group, Physical Chemistry Division, CSIR-National Chemical Laboratory, Pune 411\,008, India}};\\
$^{b}${\em{Department of Chemistry, Indian Institute of Technology Bombay, Powai, Mumbai 400\,076, India}};\\
$^{c}${\em{Theoretical Chemistry Section, Bhabha Atomic Research Centre, Trombay, Mumbai 400\,085, India}}
}

\begin{abstract}
The scalar-pseudoscalar interaction constant of PbF in its ground state electronic configuration is calculated
using the Z-vector method in the relativistic coupled-cluster framework.
The precise calculated value is very important to set upper bound limit on ${\mathcal{P,T}}$-odd scalar-pseudoscalar interaction
constant, $k_s$, from the experimentally observed ${\mathcal{P,T}}$-odd frequency shift. Further, the ratio of the effective electric
field to the scalar-pseudoscalar interaction constant is also calculated which is required to get
an independent upper bound limit of electric dipole moment of electron, $d_e$, and $k_s$ and how these ($d_e$ and $k_s$) 
are interrelated is also presented here.
\end{abstract}
\maketitle


\section{Introduction}\label{intro}
One of the biggest mysteries of our universe is the dominance of matter over the antimatter \cite{dine_2003}. The combination
of charge conjugation (${\mathcal{C}}$) and parity (${\mathcal{P}}$) symmetries (${\mathcal{CP}}$) violating
interaction along with other factors can explain this matter-antimatter asymmetry \cite{sakharov_1967}. However, the ${\mathcal{CP}}$
violation within the standard model (SM) of electroweak
and strong interaction (arising from the complex quark mixing Kobayashi-Maskawa matrix) is not strong enough to explain this
asymmetry \cite{gavela_1994}. Despite of the fact that the SM has some well known unresolved problems and drawbacks, there are very little
experimental data available which can directly contradict the SM. On the other hand, there are many extensions of the SM which can
resolve the flaws of the SM but none of them are established experimentally \cite{commins_1999, bernreuther_1991}.
\par
The electric dipole moment of electron (eEDM) arises due to the violation of  both ${\mathcal{P}}$ and time reversal
invariance (${\mathcal{T}}$) symmetries \cite{commins_1999, bernreuther_1991}.
Therefore, the eEDM experiment can explore ``new physics'' beyond the 
conventional SM \cite{fortson_2003}. According to the SM, the eEDM is too small ($d_e$ $<$ 10$^{-38}$ e cm) to be
observed experimentally \cite{khriplovich_2011}.
On the other hand, many extensions of the SM suggest that it would lie in the limit of current experimental sensitivity \cite{bernreuther_1991}.
However, the smallness in the value of eEDM restricts us to do experiments with single electron as the highest external
electric field generated in the laboratory is not large enough to observe the eEDM effect. On the other hand, diatomic
molecules have been proposed \cite{sushkov_1978, flambaum_1976} and experimented \cite{ybf_edm, tho_edm} as they offer very
high sensitivity to the EDM effect.
There are two main possible sources of permanent molecular EDM (arises only when both ${\mathcal{T}}$ and ${\mathcal{P}}$
symmetries are broken) of a paramagnetic molecule:
(i) the eEDM and (ii) the scalar-pseudoscalar (S-PS) interaction of nucleon and electron. Although the former has been studied extensively,
the latter got a little attention.
\par
\begin{figure}[ht]
\centering
\begin{center}
\includegraphics[scale=0.1, height=3.5cm]{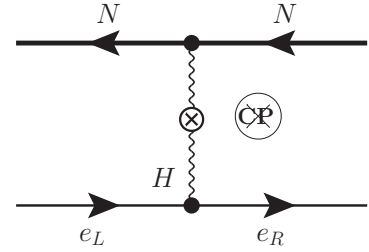}
\caption{Scalar-pseudoscalar interaction of nucleon (N) and electron mediated by an exchange of Higgs (H) particle. Here, e$_R$ and e$_L$ are
         right and left handed electrons, respectively.}
\label{sps_higgs}
\end{center}  
\end{figure}
The S-PS electron-nucleon interaction arises due to the coupling interaction between the scalar-hadronic current and the pseudoscalar
electronic current. The scalar and pseudoscalar components of neutral Higgs boson (H) particle can mediate this
interaction (Fig. \ref{sps_higgs}) \cite{barr_1992}. There is only one Higgs particle in the SM which forbids such interaction but a number
of various multi Higgs models [including the minimal supersymmetric standard model (MSSM)] predicts such interaction.
An interesting characteristic of these models is that they also predict the baryon number violation originating from
the exchange of neutral Higgs particle \cite{kazarian_1992} which is one of the other conditions to solve the matter-antimatter asymmetry of our universe.
\par
Recently, we have calculated the effective electric field ($E_\mathrm{eff}$) experienced by the unpaired electron of PbF
in its ground electronic state ($^{2}\Pi_{\frac{1}{2}}$) \cite{sasmal_pbf}. PbF has some interesting characteristics
that make it an important player for the eEDM experiment.
It offers very high $E_\mathrm{eff}$ \cite{sasmal_pbf, skripnikov_2014}.
It has a small g-factor \cite{skripnikov_gfac}, which make it very insensitive to the background magnetic field and thus many systematic errors can be
suppressed in the eEDM experiment \cite{shaferray_2006}. Two opposite parity levels of PbF are almost degenerate which suggests that it
can be polarized very easily (i.e., by weak external electric field) \cite{mcraven_2008}. It is worth to remember that in an eEDM experiment,
the molecule must be polarized completely (with minimal applied electric field) to fully utilize the $E_\mathrm{eff}$.
In this work, we have calculated the ${\mathcal{P,T}}$-odd S-PS interaction constant ($W_\mathrm{s}$) of PbF which is an important quantity
to set upper bound limit on the S-PS interaction constant ($k_s$). We have used the Dirac-Hartree-Fock (DHF) method to include
the effect of special relativity. The coupled cluster (CC) method \cite{cizek_1966, cizek_1967, bartlett_1978} is used for the correlation
treatment of opposite spin electrons.
The Z-vector method \cite{schafer_1984, zvector_1989} in the relativistic CC framework \cite{sasmal_srf} is used to calculate the ground
state properties as it can produce very accurate wavefunction in the nuclear region of PbF \cite{sasmal_pbf}.
The accuracy of the wavefunction in the nuclear region of heavy nucleus is very important for the
precise calculation of these types of ${\mathcal{P,T}}$-odd properties.
\par
The paper is organized as follows. A brief description of the scalar-pseudoscalar interaction, including an overview of the Z-vector method
are given in Sec. \ref{theory}. We present our computational details in Sec. \ref{comp}, before showing our results and discussion about those
in Sec. \ref{res}. Finally, we conclude our remarks in Sec. \ref{conc}. We have used atomic units consistently unless stated explicitly.
\section{Theory}\label{theory}
\subsection{Scalar-pseudoscalar interaction}
The pseudoscalar electronic current interacts with the scalar hadronic current and generates the S-PS electron-nucleon interaction.
The matrix element of ${\mathcal{P,T}}$-odd S-PS interaction constant, $W_\mathrm{s}$, is given by the following expression:
\begin{eqnarray}
 W_{\text{s}}=\frac{1}{\Omega k_\text{s}}\langle \Psi_{\Omega}|\sum_j^n H_{\text{SP}}(j)| \Psi_{\Omega} \rangle.
\label{W_s}
\end{eqnarray}
Here, $\Psi_{\Omega}$ is the electronic wavefunction of the $\Omega$ state where $\Omega$ is the projected value of the total angular momentum
along the molecular axis.
The dimensionless S-PS electron-nucleus coupling constant, $k_s$ can be expressed as
$k_s$=$k_{s,p}$+$(\frac{N}{Z})k_{s,n}$, where $k_{s,n}$ and $k_{s,p}$ are electron-neutron and electron-proton coupling
constant, respectively, Z and N are the number of proton and neutron in the nucleus, respectively.
$H_{\text{SP}}$ is the corresponding interaction Hamiltonian and can be given by \cite{hunter_1991}
\begin{eqnarray}
H_{\text{SP}}= i\frac{G_{F}}{\sqrt{2}}Zk_{s} \gamma^0 \gamma^5 \rho_N(r) ,
\label{H_SP}
\end{eqnarray}
where, G$_F$ is the Fermi constant, $\gamma$ are the usual Dirac matrices and $\rho_N(r)$ is the nuclear charge density normalized to unity.
\begin{table*}[ht]
\caption{ Cutoff used and correlation energy of the ground state of PbF in different basis sets}
\newcommand{\mc}[3]{\multicolumn{#1}{#2}{#3}}
\begin{center}
\begin{tabular}{lccccrcccr}
\hline
\hline
\mc{4}{c}{Basis} & \mc{2}{c}{Cutoff (a.u.)} & \mc{2}{c}{Spinor} & \mc{2}{c}{Correlation Energy (a.u.)}\\
\cline{1-4} \cline{5-6} \cline{7-8} \cline{9-10}
Name & Nature & Pb & F & Occupied & Virtual & Occupied & Virtual & MBPT(2) & CCSD \\
\hline
A & TZ & dyall.cv3z & cc-pCVTZ & -100.0  & 1000.0 & 73 & 367 & -2.67295119 & -2.43899139 \\
B & TZ & dyall.cv3z & cc-pCVTZ &    ×    & 1000.0 & 91 & 367 & -3.25352488 & -3.00168924 \\
C & QZ & dyall.cv4z & cc-pCVQZ & -100.0  &   70.0 & 73 & 449 & -2.30205292 & -2.07928886 \\
D & QZ & dyall.cv4z & cc-pCVQZ &    ×    &   70.0 & 91 & 449 & -2.36706015 & -2.14007942 \\
E & QZ(core) & dyall.ae4z+core & cc-pCVQZ & × & 40.0 & 91 & 443 & -2.55636058 & -2.31517747 \\
\hline
\hline
\end{tabular}
\end{center}
\label{basis}
\end{table*}
\subsection{Z-vector method}
The Dirac-Coulomb (DC) Hamiltonian is used to treat the relativistic motion of electrons. The DC Hamiltonian is given as
\begin{eqnarray}
{H_{DC}} &=&\sum_{i} \Big [-c (\vec {\alpha}\cdot \vec {\nabla})_i + (\beta -{\mathbb{1}_4}) c^{2} + V^{nuc}(r_i)+ \nonumber\\
       && \sum_{j>i} \frac{1}{r_{ij}} {\mathbb{1}_4}\Big],
\end{eqnarray}
where, $c$ is the speed of light, {\bf$\alpha$} and $\beta$ are the usual Dirac matrices,
${\mathbb{1}_4}$ is the 4$\times$4 identity matrix and
$V^{nuc}(r_i)$ is the nuclear potential function.
The Dirac-Hartree-Fock (DHF) method is used to solve the DC Hamiltonian and the corresponding DHF wavefunction
is used as a reference function for the correlation calculation.
As the DC Hamiltonian has unbounded solutions, the relativistic calculations are done using no-pair approximation \cite{sucher_1980, almoukhalalati_2016}.
This means that the DC Hamiltonian is nested by the projectors, which remove the negative energy solutions. DHF calculations
are performed with the implicit use of these projectors and only the positive energy orbitals are incorporated in the correlation calculations.
However, how to go beyond the no-pair approximation by accounting for correlation
contributions of negative energy states has been discussed in depth in Ref. \cite{liu_2013, liu_2014, liu_2016}.
The DHF misses the instantaneous interaction
of opposite spin electrons and we have used the coupled-cluster method to incorporate the missing dynamic electron correlation.
\par
The calculation of the ${\mathcal{P,T}}$-odd property as described in Eq. \ref{W_s} needs a very accurate wavefunction in
the nuclear region of the diatom and we have used the Z-vector method in the relativistic CC framework as it fulfills that requirement
for the ground state ($^{2}\Pi_{\frac{1}{2}}$) of PbF \cite{sasmal_pbf}.
Computationally, Z-vector method \cite{schafer_1984, zvector_1989} is a four step process: (i) calculation of the excitation operator (T),
(ii) calculation of the intermediate matrix elements using one-electron integrals, two-electron integrals and amplitudes of the excitation operator, T,
(iii) calculation of the deexcitation operator ($\Lambda$), and (iv) finally, calculation of the
desired property using the corresponding property integrals and amplitudes of T and $\Lambda$ operators.
However, the second step is optional but it can save enormous computational time.
It is worth to remember that
both T and $\Lambda$ are perturbation independent. So, we need only one set of CC calculation to obtain any number
of desired properties.
\par
The form of the coupled-cluster excitation operator, T,  is given as
\begin{eqnarray}
 T=T_1+T_2+\dots +T_N=\sum_n^N T_n ,
\end{eqnarray}
with
\begin{eqnarray}
 T_m= \frac{1}{(m!)^2} \sum_{ij\dots ab \dots} t_{ij \dots}^{ab \dots}{a_a^{\dagger}a_b^{\dagger} \dots a_j a_i} ,
\end{eqnarray}
where, i,j are the hole and a,b are the particle indices and $t_{ij..}^{ab..}$ are the cluster amplitudes corresponding 
to the cluster operator $T_m$.
The coupled-cluster wavefunction is given by
\begin{eqnarray}
|\Psi_{cc}\rangle=e^{T}|\Phi_0\rangle ,
\end{eqnarray}
where, $\Phi_0$ is the DHF wavefunction.
In the coupled-cluster single and double (CCSD) model, $T=T_1+T_2$. The equations for T$_1$ and T$_2$ can be given as
\begin{eqnarray}
 \langle \Phi_{i}^{a} | (H_Ne^T)_c | \Phi_0 \rangle = 0 , \,\,
  \langle \Phi_{ij}^{ab} | (H_Ne^T)_c | \Phi_0 \rangle = 0 ,
 \label{cc_amplitudes}
\end{eqnarray}
where, H$_N$ is the normal ordered DC Hamiltonian and subscript $c$ means only the connected terms exist in the
contraction between H$_N$ and T. This connectedness ensures the size-extensivity.
\par
The form of the deexcitation operator, $\Lambda$, is given as
\begin{eqnarray}
 \Lambda=\Lambda_1+\Lambda_2+ \dots+\Lambda_N=\sum_n^N \Lambda_n ,
\end{eqnarray}
with
\begin{eqnarray}
 \Lambda_m= \frac{1}{(m!)^2} \sum_{ij \dots ab \dots} \lambda_{ab \dots}^{ij \dots}{a_i^{\dagger}a_j^{\dagger} \dots a_b a_a} ,
\end{eqnarray}
where, $\lambda_{ab \dots}^{ij \dots}$ are the cluster amplitudes corresponding  to the operator $\Lambda_m$.
In the CCSD model, $\Lambda=\Lambda_1+\Lambda_2$. The equations for the amplitudes of $\Lambda_1$
and $\Lambda_2$ operators are given by \cite{zvector_1989}
\begin{eqnarray}
\langle \Phi_0 |[\Lambda (H_Ne^T)_c]_c | \Phi_{i}^{a} \rangle + \langle \Phi_0 | (H_Ne^T)_c | \Phi_{i}^{a} \rangle = 0,
\end{eqnarray}
\begin{eqnarray}
\langle \Phi_0 |[\Lambda (H_Ne^T)_c]_c | \Phi_{ij}^{ab} \rangle + \langle \Phi_0 | (H_Ne^T)_c | \Phi_{ij}^{ab} \rangle \nonumber \\
 + \langle \Phi_0 | (H_Ne^T)_c | \Phi_{i}^{a} \rangle \langle \Phi_{i}^{a} | \Lambda | \Phi_{ij}^{ab} \rangle = 0.
\label{lambda_2}
\end{eqnarray}
Finally, the energy derivative can be obtained as
\begin{eqnarray}
 \Delta E' = \langle \Phi_0 | (O_Ne^T)_c | \Phi_0 \rangle + \langle \Phi_0 | [\Lambda (O_Ne^T)_c]_c | \Phi_0 \rangle
\end{eqnarray}
where, $O_N$ is the normal ordered one-electron property operator.
\section{Computational details}\label{comp}
In this work, we have used the DIRAC10 program package \cite{dirac10} to solve the DHF equation and to generate one-, and two-body matrix elements.
The property integrals are constructed by using a locally modified version of DIRAC10 program package.
Large and small components basis are linked through restricted kinetic balance condition \cite{dyall_2007}.
The basis functions are expressed in scalar basis and all unphysical solutions are removed by means of the
diagonalization of free particle Hamiltonian. This generates equal number of positronic and electronic orbitals.
We have used the finite size of nucleus where the Gaussian charge distribution is considered.
The exponents for the nuclear parameters are taken as default values of DIRAC10 \cite{visscher_1997}.
We have used two sets of basis - one with triple zeta (TZ) basis (dyall.cv3z for Pb \cite{dyall_4-6_p} and cc-pCVTZ
for F \cite{ccpcvxz_h_b-ne}) and the other with
quadruple zeta (QZ) basis (dyall.cv4z for Pb \cite{dyall_4-6_p} and cc-pCVQZ for F \cite{ccpcvxz_h_b-ne}).
In each set of basis, we have done two calculations- one with 73 and another with 91 correlated electrons.
The virtual orbital cutoff used for TZ and QZ calculations are 1000 a.u. and 70 a.u., respectively.
We have done one more all electron calculation using QZ basis with the explicit use of core correlating functions,
which are taken from Ref. \cite{dyall_core_2012} (given in Appendix A).
The cutoff used for this calculation is 40 a.u.
The experimental bond length of PbF (3.89 a.u.) \cite{herzberg_4} is used in all the calculations.
\section{Results and discussion}\label{res}
The aim of the present study is to provide the accurate value of S-PS interaction constant,
$W_\mathrm{s}$, as it is an important quantity to set upper bound limit on the ${\mathcal{P,T}}$-odd S-PS interaction
constant, $k_s$, combined with the experimentally observed ${\mathcal{P,T}}$-odd frequency change.
The ratio of $E_\mathrm{eff}$ to $W_\mathrm{s}$ is also calculated here as it provides the interrelation between $d_e$ and $k_s$.
Previously, in Ref. \cite{sasmal_pbf}, we have calculated the $E_\mathrm{eff}$ of PbF using the Z-vector method in the relativistic coupled-cluster
framework. In that paper, we also showed that the calculated value of $E_\mathrm{eff}$ is very reliable as the wavefunction in the
nuclear region is very accurate which is evident from the calculated value of the parallel component of magnetic hyperfine
structure constant of PbF. So, in this work,
we have opted the same basis and cutoff for the calculation of $W_\mathrm{s}$ in the ground state of PbF.
In Table \ref{basis}, we present the basis, cutoff and the correlation energies of PbF.
\begin{table}[ht]
\caption{$E_\mathrm{eff}$ (in GV/cm), $W_\mathrm{s}$ (in kHz) and the ratio of them (R = $E_\mathrm{eff}$/$W_\mathrm{s}$ in units of 10$^{18}$/e cm) of PbF.}
\newcommand{\mc}[3]{\multicolumn{#1}{#2}{#3}}
\begin{center}
\begin{tabular}{lclcccc}
\hline
\hline
Basis & \mc{2}{c}{$E_\mathrm{eff}$} &  \mc{2}{c}{$W_\mathrm{s}$}  & \mc{2}{c}{R}\\
\cline{2-3}  \cline{4-5}  \cline{6-7}
× & SCF & Z-vector & SCF & Z-vector & SCF & Z-vector\\
\hline
A(TZ,73e) & 39.8 & 37.5 \cite {sasmal_pbf} & 91.1 & 86.6 & 105.6 & 104.7 \\
B(TZ,91e) & 39.8 & 38.1                    & 91.1 & 88.2 & 105.6 & 104.4 \\
C(QZ,73e) & 39.6 & 37.9 \cite {sasmal_pbf} & 90.6 & 87.8 & 105.7 & 104.4 \\
D(QZ,91e) & 39.6 & 38.1 \cite {sasmal_pbf} & 90.6 & 88.1 & 105.7 & 104.6 \\
E[QZ(core),91e] & 39.6 & 38.2              & 90.6 & 88.4 & 105.7 & 104.5 \\
\hline
\hline
\end{tabular}
\end{center}
\label{pbf_pt}
\end{table}
\begin{table*}[ht]
\caption{ Comparison of $W_\mathrm{s}$ of the ground state of PbF in different methods}
\newcommand{\mc}[3]{\multicolumn{#1}{#2}{#3}}
\begin{center}
\begin{tabular}{lcc}
\hline
\hline
Method & Reference & $W_\mathrm{s}$ (kHz)\\
\hline
SODCI(13e) & Baklanov {\it et al} \cite{baklanov_2010} & 75 \\
SODCI(13e)+OC & Baklanov \cite{baklanov_thesis} & 83 \\
2c-CCSD(31e) & Skripnikov {\it et al} \cite{skripnikov_2014} & 93 \\
2c-CCSD(T)(31e) & Skripnikov {\it et al} \cite{skripnikov_2014} & 91 \\
4c-Z-vector(QZ, all electron) & This work (basis D) & 88.1\\
4c-Z-vector(QZ(core), all electron) & This work (basis E) & 88.4\\
\hline
\hline
\end{tabular}
\end{center}
\label{comparison}
\end{table*}
\par
The calculated values of $W_\mathrm{s}$ in different basis and cutoff are presented in Table \ref{pbf_pt}.
In Ref. \cite{sasmal_pbf}, it was shown that the all electron calculation QZ basis produces very precise wavefunction in the near
nuclear region of PbF. To see the core consistency of the basis, we have done one calculation in QZ basis with core correlating
functions (basis E) and the obtained value of magnetic hyperfine structure constant is 10160 MHz. The corresponding
experimental value is 10147 MHz \cite{petrov_2013, mawhorter_2011}. This shows that the 
most reliable value of $W_\mathrm{s}$ is 88.4 kHz as the all electron (91e) calculation in core consistent QZ basis
produces the most accurate wavefunction in the nuclear region of PbF.
This high value of $W_\mathrm{s}$ in PbF shows that S-PS interaction can contribute a significant amount to the
total molecular EDM of PbF.
\par
We have also compared our Z-vector values of $W_\mathrm{s}$ with other available theoretical results.
Baklanov {\it et al} \cite{baklanov_2010} calculated the value of $W_\mathrm{s}$ as 75 kHz using spin-orbit direct CI (SODCI) method
but without outer core (OC) correction. Later, they included the OC correction and improved their results by
8 kHz \cite{baklanov_thesis}. However, the truncated CI does not scale properly with the number of electrons as it is not size-extensive.
Thus, for the calculation of heavy-atom containing system where the number of electrons is significant,
coupled-cluster may be a better alternative than the truncated CI method.
The two-component (2c) relativistic CC calculations are done by Skripnikov {\it et al} \cite{skripnikov_2014} using CCSD approximation and obtained
the value of $W_\mathrm{s}$ as 93 kHz. Their partial triples correction in the CCSD (CCSD(T)) calculation decreases
the value of $W_\mathrm{s}$ by 2 units \cite{skripnikov_2014}. Skripnikov {\it et al} used the ``valence'' semilocal version of the
GRECP scheme \cite{titov_1999}
in their calculation and correlated only 31 electrons explicitly. The valence GRECP calculation can introduce significant errors
in the valence electronic state if the nuclear screening effects are not properly reproduced \cite{mosyagin_2010}.
However, in Ref. \cite{sasmal_pbf}, we have
shown that the explicit treatment of all electrons is very important for this types of ``atom in compound'' \cite{AIC} properties.
That's why in this work, we have done all electron correlation calculations for both TZ and QZ basis which signifies the
reliability of our calculated results.
\par
The $E_\mathrm{eff}$ values presented in Table \ref{pbf_pt}
are taken from Ref. \cite{sasmal_pbf} except for the basis B and E which have been calculated here
explicitly. The ratio ($R$) of $E_\mathrm{eff}$ to $W_\mathrm{s}$ is also calculated and presented in the same table
as it is a very important quantity to decouple the eEDM and S-PS effects using two different eEDM experiments.
Dzuba {\it et al} \cite{dzuba_2011} suggested that this ratio is a characteristic of the heavy nucleus, i.e., it would
be more or less same for a specific
heavy nucleus independent of the diatom. This is because of the following reasons: (i) these types of
${\mathcal{P,T}}$-odd properties are predominantly dependent on the amplitude of the wavefunction in the nuclear region and for
a specific angular momentum, the Dirac equation becomes same for every single-electron states in that small distance;
(ii) the many-body effects like core polarization, etc have insignificant effects on the ratio ($R$) as the predominant contribution
comes from the valance $s_{1/2}-p_{1/2}$ matrix elements.
From our Z-vector and self-consistent field (SCF) values, we can see that the correlation treatment changes the value of $R$
only by $\sim$ 1\% which supports the previous argument of Dzuba {\it et al} \cite{dzuba_2011}.
The most reliable value of $R$ is $104.5 \times 10^{18} e^{-1} cm^{-1}$
which is calculated using QZ basis by correlating all electrons.
Dzuba {\it et al} also calculated the value of $R$ for $^{208}$Pb nucleus by using analytical expressions
to calculate the $s_{1/2}-p_{1/2}$ matrix element of the corresponding operators of $R$ \cite{dzuba_2011}. Our values are very close
to their value of $R$ ($111.6 \times 10^{18} e^{-1} cm^{-1}$).
Using this value of $R$, the interrelation of $d_e$ and $k_s$ becomes (for details see Ref. \cite{sasmal_hgh})
\begin{equation}
 d_e + 4.78 \times 10^{-21} k_s = d_e^\mathrm{expt}|_{\!_{k_s=0}} ,
 \label{relation}
\end{equation}
where, $d_e^\mathrm{expt}|_{\!_{k_s=0}}$ is the eEDM limit derived from the ${\mathcal{P,T}}$-odd frequency change of PbF experiment
at the limit of $k_s$ = 0. \par
From Table \ref{pbf_pt}, it can be seen that there is a change of the order of 0.5\% in the property values due to the further
augmentation of the basis set. Therefore, it is expected that a saturated core-basis set might lead to further
changes of the same order of magnitude. We, therefore, expect a similar accuracy of 4\% for the calculation of
$W_\mathrm{s}$ of PbF, since used basis set and the cutoff is same, what has been used in Ref. \cite{sasmal_pbf}. 
\section{Concluding remarks}\label{conc}
In summary, we have performed the Z-vector calculation in the relativistic coupled-cluster framework to obtain the
S-PS interaction constant, $W_\mathrm{s}$, in the ground state of PbF. We have also calculated the ratio of
$E_\mathrm{eff}$ to $W_\mathrm{s}$ to get a relation between $d_e$ and $k_s$ which in turn can help us to
get model independent limit of $d_e$ and $k_s$.
\section*{Acknowledgements}
Authors acknowledge a grant from CSIR 12th Five Year Plan project on Multi-Scale Simulations of Material (MSM)
and the resources of the Center of Excellence in Scientific Computing at CSIR-NCL. S.S. and K.T. acknowledge the CSIR
for their fellowship.
S.P. acknowledges funding from J. C. Bose Fellowship grant of Department of Science and Technology (India).
\par
This paper is dedicated to Professor Debashis Mukherjee on the occasion of his 70$^{th}$ birthday.

\begin{widetext}
\section*{Appendix A: Core correcting functions for QZ basis of Pb} \label{core_pb}
\begin{table*}[ht]
\caption{Core correcting functions for QZ basis of Pb}
\newcommand{\mc}[3]{\multicolumn{#1}{#2}{#3}}
\begin{center}
\begin{tabular}{lcccccc}
\hline
\hline
Shell & \mc{6}{c}{Functions} \\
\hline
5d & f & f & f & g & g & h\\
× & 2.8528652E+00 & 1.3451403E+00 & 5.7943678E-01 & 2.3827416E+00 & 1.0085055E+00 & 1.8714871E+00\\
4f & g & g & g & h & h & i\\
× & 3.8173731E+01 & 1.4553553E+01 & 5.6721782E+00 & 2.6860715E+01 & 9.9334269E+00 & 1.7969506E+01\\
3d & g & g & h & × & × & ×\\
× & 1.6959303E+02 & 6.5883462E+01 & 1.1822359E+02 & × & × & ×\\
2s2p & g & × & × & × & × & ×\\
× & 6.3901836E+02 & × & × & × & × & ×\\
1s & f & × & × & × & × & ×\\
× & 9.2423036E+03 & × & × & × & × & × \\
\hline
\hline
\end{tabular}
\end{center}
\label{core_fun}
\end{table*}
\end{widetext}
\end{document}